\documentclass{PoS}
\RequirePackage{xspace}
\usepackage{relsize}

\def\babar{\mbox{\slshape B\kern-0.1em{\smaller A}\kern-0.1em
    B\kern-0.1em{\smaller A\kern-0.2em R}}}
\def\Kbar    {\kern 0.18em\overline{\kern -0.18em K}{}\xspace}

\def\Kz      {\ensuremath{K^0}\xspace}
\def\Kzb     {\ensuremath{\Kbar^0}\xspace}
\def\KzKzb   {\ensuremath{\Kz {\kern -0.16em \Kzb}}\xspace}

\def\Ks     {\ensuremath{K_S}\xspace}
\def\Kl     {\ensuremath{K_L}\xspace}
\def\KsKs   {\ensuremath{\Ks {\kern -0.16em \Ks}}\xspace}
\def\KlKl   {\ensuremath{\Kl {\kern -0.16em \Kl}}\xspace}
\def\KsKl   {\ensuremath{\Ks {\kern -0.16em \Kl}}\xspace}
\def\KlKs   {\ensuremath{\Kl {\kern -0.16em \Ks}}\xspace}
\def\Dbar    {\kern 0.18em\overline{\kern -0.18em D}{}\xspace}
\def\DDb   {\ensuremath{D {\kern 0.05em \Dbar}}\xspace}
\def\Dz      {\ensuremath{D^0}\xspace}
\def\Dzb     {\ensuremath{\Dbar^0}\xspace}
\def\DzDzb   {\ensuremath{\Dz {\kern 0.05em \Dzb}}\xspace}
\def\Bbar    {\kern 0.18em\overline{\kern -0.18em B}{}\xspace}

\def\Bz      {\ensuremath{B^0}\xspace}
\def\Bzb     {\ensuremath{\Bbar^0}\xspace}
\def\BzBzb   {\ensuremath{\Bz {\kern -0.16em \Bzb}}\xspace}
\def\Bu      {\ensuremath{B^+}\xspace}
\def\Bub     {\ensuremath{B^-}\xspace}

\def\BpBm    {\ensuremath{\Bu {\kern -0.16em \Bub}}\xspace}

\def\Bsb     {\ensuremath{\Bbar_s}\xspace}
\def\lplm     {\ensuremath{l^+ {\kern 0.05em l^-}}\xspace}
\def\ee     {\ensuremath{e^+ {\kern 0.05em e^-}}\xspace}
\def\mm     {\ensuremath{\mu^+ {\kern 0.05em \mu^-}}\xspace}
\def\nub    {\kern 0.18em\overline{\kern -0.18em \nu}{}\xspace}
\def\nunu   {\ensuremath{\nu {\kern 0.05em \nub}}\xspace}

\title{Rare and forbidden decay of charm mesons at CLEO-c and BES-III }

\ShortTitle{Rare and forbidden decay of charm mesons at CLEO-c and
BES-III}

\author{\speaker{Hai-Bo Li}\thanks{For BES-III and CLEO-c Collaborations.}\\
        Author Institute of High Energy Physics,  P.O.Box 918, Beijing  100049, China \\
        E-mail: \email{lihb@ihep.ac.cn}}

\abstract{This paper presents the results of searches for the
flavor-changing neutral current, lepton-number-violating and
lepton-flavor-violating processes in the charm mesons decays.
Recent results from charged $D$ mesons decays to final states with
dielectrons from CLEO-c are reported. The prospects of the
searches for the rare charm decays at BES-III are also discussed.}

\FullConference{KAON International Conference\\
         May 21-25 2007\\
         Laboratori Nazionali di Frascati dell'INFN, Rome, Italy}

\begin{document}

\section{Introduction}

Searches for rare-decay processes have played an important role in
the development of the Standard Model (SM).  The absence of
flavor-changing neutral current (FCNC) in $K$ decays implied the
existence of charm quark, and the observations of $B^0 - \Bzb$ and
$B_s - \Bsb$ mixings (FCNC processes) signaled the very large
top-quark mass. The study of the FCNC has been focused on rare
decays involving transitions such as $s \rightarrow d \lplm$, $s
\rightarrow d \nunu$, $b \rightarrow s \gamma$, and $b \rightarrow
\lplm$.

In contrast with the $K$ and $B$ FCNC processes, the $D$ meson
FCNC transitions are mediated by light down-quark sector, which
implies an efficient Glashow-Iliopoulos-Maiani (IM) cancellation.
 In this case, $D^0$ mixing, as well as FCNC decays are expected to
be very small in the SM short-distance contributions.  Many
extensions of the SM may enhance the mixing rate and FCNC
processes in the short-distance of the $D$ system by orders of
magnitudes. Unfortunately, the new physics contributions are
diluted by large long-distance contributions which are likely to
dominate over the SM short-distance effects.  For example, the
charm radiative decays $D \rightarrow h \gamma$ are completely
dominated by hadronic uncertainty, where $h$ denotes a light
hadron final state~\cite{ian_2003}. However, semileponic decays
such as $c \rightarrow u \lplm$ may be used to constrain the new
physics since one can look at the new physics contributions to the
whole kinematics to which the new physics may contribute the
region away from the resonance-dominated (long-distance dominated)
region~\cite{Burdman_2002,fajfer_2001}.
 Although the long-distance effects dominate the $D^0$ mixing
 rate, the recent experimental bounds can still be used to
 constrain the new physics space. Purely leptonic flavor-violating
 (LFV) modes such as $D^0 \rightarrow \mu^\pm e^\mp$ and $D \rightarrow
 h \mu^\pm e^\mp$, as well as lepton-number violated (LNV) modes $D^+
\rightarrow h^- e^+e^+$ ,  are completely not allowed in the SM,
they are  "smoking gun" for new physics searches.  In this paper,
we report
 the recent measurements of rare and forbidden charm decays from CLEO-c, and
 discuss the prospect of these decays at the BES-III experiment.
 A comparison between $B$ factories and $\tau$-charm factories are
 also presented.

\section{CLEO-c and BES-III for charm flavor physics}

 The Cornell Electron Storage Ring (CESR) had been upgraded to
 CESR-c with the installation of 12 wiggler magnets to increase
 damping at low energy. The CLEO-c detector is minimal
 modification of the well understood CLEO-III detector. It is the
 first modern detector to operate at charm threshold. The CLEO-c
 has accumulated a total of 281 pb$^{-1}$ at $\psi(3770)$ peak (1.8 million
 $\DDb$ pairs) and 200 pb$^{-1}$ at $\sqrt{s} = 4170$ MeV for
 $D_s$ physics. CLEO-c expects to take data until April 2008 and
 will approximately triple each data set by that time~\cite{ian_2006}.

 The BES-III detector is designed for the
$e^{+}e^{-}$ collider running at the $\tau$-Charm energy region,
called BEPCII, which is currently under construction at IHEP,
Beijing, P.R. China~\cite{wang_2006}. The accelerator has two
storage rings with a circumstance of 224~m, one for electron and
one for positron, each with 93 bunches spaced by 8~ns. The total
current of the beam is 0.93~amp, and the crossing angle of two
beams is designed to be 22~mrad. The peak luminosity is expected
to be $10^{33} \mbox{cm}^{-2}\mbox{s}^{-1}$ at the beam energy of
1.89 GeV,  the bunch length is estimated to be 1.5~cm and the
energy spread will be $5.16\times 10^{-4}$. At this moment, the
LINAC has been installed and successfully tested, all the
specifications are satisfied. The storage rings have installed,
and will be commissioned for synchrotron radiation run by the end
of the year.

The BES-III detector consists of a He-based small cell drift
chamber, Time-Of-Flight (TOF) counters for PID, a CsI(Tl) crystal
calorimeter, a solenoid super-conducting magnet with a field of 1
Tesla and the magnet yoke interleaved with Resistive Plate
Chambers (RPC)  counters as the muon chamber. The construction is
expected to be completed in the middle of 2007. Photon energy
resolution is $\Delta E/E = 2.5\%$ at $E_{\gamma}= 1.0$ GeV. The
momentum resolution is $\sigma_p/p = 0.5\%$ at $p= 1.0$ GeV/$c$,
and the $dE/dx$ resolution for hadron tracks is about 6\%. The
time resolution of TOF is about 100 ps, combining energy loss
($dE/dx$) measurement in the draft chamber, give 10 $\sigma$
K/$\pi$ resolution across the typical kinematic range.

\begin{table}[htbp]
\caption{$\tau$-Charm productions at BEPC-II in one year's
running($10^7s$).} \center{\begin{tabular}{@{}llll}
 \hline
               & Central-of-Mass energy & Luminosity     & \#Events  \\
Data Sample    & (MeV)& ($10^{33}$cm$^{-2}$s$^{-1}$)  & per year  \\
\hline
$J/\psi$ &  3097  & 0.6     & $10\times 10^9$\\
$\tau^+\tau^-$   & 3670 & 1.0  & $12\times 10^6$ \\
$\psi(2S)$ & 3686  & 1.0 & $3.0\times 10^9$ \\
$D^0\overline{D}^0$ & 3770 &1.0 & $18\times 10^6$ \\
$D^+D^-$ & 3770  &1.0 & $14\times 10^6$ \\
$D^+_S D^-_S$ & 4030  &0.6 & $1.0\times 10^6$ \\
$D^+_S D^-_S$ & 4170  &0.6 & $2.0\times 10^6$ \\
\hline
\end{tabular} \label{tab:lum}}
\end{table}

The BES-III can accumulate $10\times10^9$ $J/\psi$, $3\times 10^9$
$\psi(2S)$ , 30 million $D\overline{D}$ or 2 million $D_S
\overline{D}_S$-pairs per running year as listed in
Table~\ref{tab:lum}, respectively, when it is turned to run at
resonances in 2008~\cite{li_2006}. Coupled with what is available
at CLEO-c, the BES-III will make it possible for the first time to
study in detail the light hadron spectroscopy in the decays of
charmonium states and the charmed mesons. In addition, about 30
million $D\overline{D}$ pairs will be collected at BES-III in one
year at $\psi(3770)$ peak. With modern techniques and
unprecedented high statistical data sample, Searching for rare D,
Charmonium and tau decays will be possible, such as LFV and LNV
decays, invisible decays.

\section{Rare charm decays from CLEO-c and prospect at BES-III}

With 1.6 million charged $D$ at CLEO-c, they have searched for the
FCNC decays $D^+\rightarrow \pi^+ e^+e^-$ and $D^+ \rightarrow
e^+e^-$, and the LNV decays $D^+ \rightarrow \pi^- e^+e^+$ and
$D^+ \rightarrow K^- e^+e^-$ (charge-conjugate modes are implicit
throughout this paper)~\cite{cleo-c-rare-2005}. For each candidate
decay of the form $D^+ \rightarrow h^\pm e^\mp e^+$, where $h$ is
either $\pi$ or $K$, the energy difference $\Delta E = E_{cand}
-E_{beam}$ is computed and the beam-constrained mass difference
$\Delta m_{bc} = \sqrt{E^2_{beam} - | \vec{p}_{cand}|^2} -
m_{D^+}$, where $E_{cand}$ and $\vec{p}_{cand}$ are the measured
energy and momentum of the $h^\pm e^{\mp} e^+$ candidate,
$E_{beam}$ is the beam energy, and $m_{D^+}$ is the niminal mass
of the $D^+$ meson. Event with $D^+$ candidates satisfying $-30$
MeV $\leq \Delta m_{bc} < 30 $ MeV and -100 MeV $\leq \Delta E <
100 $ MeV are selected for further study. Within this region, the
signal box is defined as $-5$ MeV $\leq \Delta m_{bc} < 5$ MeV and
-20 MeV $\leq \Delta E < 20 $ MeV,  corresponding to $\pm 3
\sigma$ in each variable, as determined by MC simulation. The
remainder of the candidate sample was used to assess backgrounds.

  Candidates for the decay of the long-distance decay $D^+
 \rightarrow \pi^+ \phi \rightarrow \pi^+ e^+e^-$ are selected
 based on the mass squared of the final-state $e^+e^-$ (equal to
 the $q^2$ of the decay), with 0.9973 GeV$^2$ $\leq m^2_{e^+e^-} <
 1.0813$ GeV$^2$ defining the $\phi$-resonant region. This region
 is used both to veto the long-distance $D^+ \rightarrow \phi
 \pi^+ \rightarrow \pi^+ e^+e^-$ contribution and to measure its
 branching fraction. Backgrounds in the $D^+ \rightarrow h^\pm
 e^\mp e^+$ candidate sample arise from both $\DDb$ and
 non-$\DDb$ sources. A detailed analysis of backgrounds can be
 found in Ref.~\cite{cleo-c-rare-2005}.

\begin{figure*}
\centering
\includegraphics[width=\textwidth]{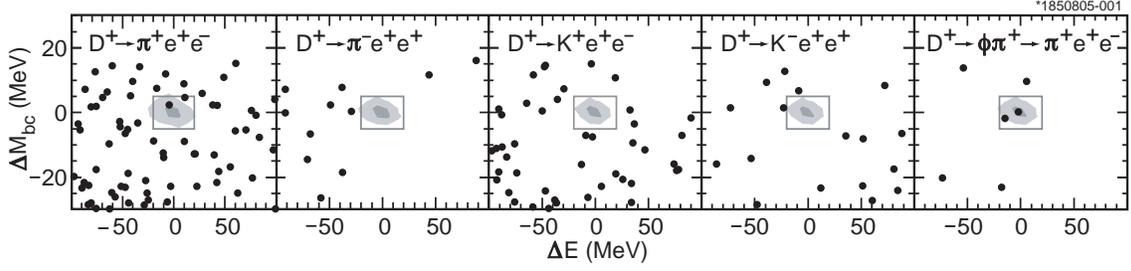}
\caption{Scatter plots of $\Delta m_{bc}$ vs.\ $\Delta E$ obtained
from data for each decay mode. The signal region, defined by $-20$
MeV $\le \Delta E < 20$ MeV and $-5$ MeV $\le \Delta m_{bc} < 5$
MeV, is shown as a box.  The two contours for each mode enclose
regions determined with signal MC simulations to contain $50\%$
and $85\%$ of signal events, respectively. } \label{cleo-c}
\end{figure*}

CLEO-c results show no evidence for signals,  the $90 \%$
confidence level upper limits are set as below:
\[
\begin{array}{lcr}
\mathcal{B}(D^+ \rightarrow \pi^+ e^+ e^-) & < & 7.4\times 10^{-6}\\
\mathcal{B}(D^+ \rightarrow \pi^- e^+ e^+) & < & 3.6\times 10^{-6}\\
\mathcal{B}(D^+ \rightarrow K^+ e^+ e^-) & < & 6.2\times 10^{-6}\\
\mathcal{B}(D^+ \rightarrow K^- e^+ e^+) & < & 4.5\times 10^{-6}
\end{array}
\]
these results for these dielectron modes are significantly more
restrictive than previous limits, and reflect sensitivity
comparable to the searches for dimuon modes ~\cite{pdg_2006}. Due
to the dominance of long-distance effects in FCNC modes, they
separately measure the branching fraction of the resonant decay
$D^+ \rightarrow \pi^+ \phi \rightarrow \pi^+ e^+ e^-$, obtaining
$\mathcal{B}(D^+ \rightarrow \phi \pi^+\rightarrow \pi^+ e^+ e^-)
= (2.7^{+3.6}_{-1.8} \pm 0.2) \times 10^{-6}$. This is consistent
with the product of known world average~\cite{pdg_2006} branching
fractions, $ \mathcal{B}(D^+ \rightarrow \phi \pi^+ \rightarrow
\pi^+ e^+ e^-) =
      \mathcal{B}(D^+ \rightarrow \phi \pi^+)
      \times
      \mathcal{B}(\phi \rightarrow e^+ e^-)
      = [(6.2\pm 0.6) \times 10^{-3}] \times [(2.98\pm 0.04) \times 10^{-4}]
      = (1.9 \pm 0.2) \times 10^{-6}.
$

\begin{table}[htbp]
\caption{Current and projected 90\%-CL upper limits on rare $D^+$
decay modes at BES-III with 20 fb$^{-1}$ data at $\psi(3770)$
peak. We assume the selection efficiencies for all modes are
35\%.} \center{\begin{tabular}{@{}llll} \hline
                & Reference     & Best Upper  &  BES-III  \\
Mode            & Experiment              & limits($10^{-6}$)      & ($\times 10^{-8}$)  \\
\hline
$ \pi^+ e^+e^-$     & CLEO-c~\cite{cleo-c-rare-2005} & 7.4 & 5.6  \\
$ \pi^+ \mu^+\mu^-$ & FOCUS~\cite{focus}  & 8.8 & 8.7  \\
$ \pi^+ \mu^+ e^-$ & E791~\cite{e791}  & 34 &  8.3 \\
$ \pi^- e^+ e^+$ & CLEO-c~\cite{cleo-c-rare-2005}  & 3.6 & 5.6  \\
$ \pi^- \mu^+\mu^+$ & FOCUS~\cite{focus}  & 4.8 & 8.7  \\
$ \pi^- \mu^+ e^+$ & E791~\cite{e791}  & 50 & 5.9 \\
$ K^+ e^+e^- $ & CLEO-c~\cite{cleo-c-rare-2005}  & 6.2 & 6.7  \\
$ K^+ \mu^+\mu^- $ & FOCUS~\cite{focus}  & 9.2 & 10.5  \\
$ K^+ \mu^+ e^- $ & E791~\cite{e791}  & 68 & 8.3 \\
$ K^- e^+ e^+ $ & CLEO-c~\cite{cleo-c-rare-2005}  & 4.5 & 6.7  \\
$ K^- \mu^+ \mu^+ $ & FOCUS~\cite{focus}  & 13 & 10.4  \\
$ K^- \mu^+ e^+ $ & E687~\cite{e687}  & 130 & 8.3  \\
\hline
\end{tabular}\label{tab:rare}}
\end{table}
\begin{table}[htbp]
\caption{Current and projected 90\%-CL upper limits on rare $D^0$
decay modes at BES-III with 20 fb$^{-1}$ data at $\psi(3770)$
peak.} \center{\begin{tabular}{@{}llll} \hline
                & Reference     & Best Upper  &  BES-III  \\
Mode            & Experiment              & limits($10^{-6}$)      & ($\times 10^{-8}$)  \\
\hline
$\gamma \gamma $  & CLEO~\cite{cleo-d0} & 28 & 5.0  \\
$\mu^+\mu^-$ & BaBar~\cite{d0-coll}  & 1.3 & 17  \\
$\mu^+ e^-$ & BaBar~\cite{d0-coll}  & 0.81 & 4.3  \\
$e^+ e^-$ & BaBar~\cite{d0-coll}  & 1.2  & 2.4  \\
$\pi^0 \mu^+\mu^-$ & E653~\cite{e653}  & 180 & 12.3  \\
$\pi^0 \mu^+ e^+$ & CLEO~\cite{cleo-d01}  & 86 & 9.7  \\
$\pi^0 e^+e^- $ & CLEO~\cite{cleo-d01}  & 45 & 7.9  \\
$K_S \mu^+\mu^- $ & E653~\cite{e653}  & 260 & 10.6  \\
$K_S \mu^+ e^- $ & CLEO~\cite{cleo-d01}  & 100 &  9.6 \\
$K_S e^+ e^- $ & CLEO~\cite{cleo-d01}  & 110 & 7.5  \\
$\eta \mu^+ \mu^- $ & CLEO~\cite{cleo-d01}  & 530 & 10  \\
$\eta \mu^+ e^- $ & CLEO~\cite{cleo-d01}  & 100 & 10  \\
$\eta e^+ e^- $ & CLEO~\cite{cleo-d01}  & 110 & 10  \\
\hline
\end{tabular}\label{tab:rare_d0} }
\end{table}

With 20 fb$^{-1}$ data at $\psi(3770)$ peak with the BES-III
detector,  the sensitivity of the measurements of the rare charm
decays are summarized for $D^+$ and $D^0$, respectively, in
Tables~\ref{tab:rare} and \ref{tab:rare_d0}. For most of them, the
sensitivities can be at the order of $10^{-8}$.  The current best
experimental limits are also listed in the tables~\ref{tab:rare}
and~\ref{tab:rare_d0}~\cite{li_2006}.

\section{$D^0-\Dzb$ Mixing, $CP$ Violation at BES-III}

With the design luminosity of $10^{33}$cm$^{-2}$s$^{-1}$, BES-III
will have the opportunity to probe for the possible new physics
which may enter up-type-quark decays. It includes searches for
charm mixing, $CP$ violation and rare charm decays. The BES-III
charm physics program also includes a variety of measurements that
will improve the determination of $\phi_3/\gamma$ from $B$-factory
experiments.  The total number of charm mesons accumulated at
BES-III will be much smaller than that at the $B$-factories which
are about 500 million $D \overline{D}$ pairs for each of them.
However, the quantum correlations in the $\psi(3770) \rightarrow
\DDb$  system will provide a unique laboratory in which to study
charm~\cite{asner_2005}.

\subsection{$D^0- \Dzb$ Mixing}

$D^0 - \Dzb$ mixing within the SM are highly suppressed due to GIM
mechanism, thus, at BES-III, searches for neutral charm mixing and
$CP$ violation in the charm decays may be essential to search for
some intriguing signals due to new physics.

The time evolution of $D^0 -\Dzb$ system, assuming no $CP$
violation in mixing, is governed by four parameters: $x=\Delta
m/\Gamma$ and $y = \Delta \Gamma/2\Gamma$ which are the mass and
width differences of
 $D$ meson mass eigenstates and characterize the mixing matrix, $\delta$
 the relative strong phase between the Cabibbo favor (CF) and the doubly-Cabibbo
suppressed (DCS) amplitudes and $R_D$ the DCS decay rate relative
to the CF decay rate. The mixing rate $R_M$ is defined as
$\frac{1}{2}(x^2+y^2)$. The SM  based predictions for $x$ and $y$,
as well as a variety of non-SM expectations, span several orders
of magnitude~\cite{ian_2003,nir_1999,petrov_2005} which is $x$
$\sim$ $y$ $\sim$ $10^{-3}$. Presently, experimental information
about charm mixing parameters $x$ and $y$ comes from the
time-dependent analyses.

At the B factories, the wrong-sign (WS) process, $D \rightarrow
K^+ \pi^-$, is used to extract the $D^0$ mixing parameters by
fitting the time-dependent decay rates.  The WS process can
proceed either through direct doubly-Cabibbo-suppressed (DCS)
decay or through mixing followed by the right-sign (RS) Cabibbo
favored (CF) decay $D^0 \rightarrow \Dzb \rightarrow K^+\pi^-$.
The two decays can be distinguished by the decay-time
distribution. For $|x|$, $y|\ll 1$, and assuming negligible $CPV$,
the decay-time distribution for $D^0 \rightarrow K^+ \pi^-$ can be
expressed as
\begin{eqnarray}
 R_{WS}(t) = e^{-\Gamma t}( R_D + \sqrt{R_D} y^{\prime} (\Gamma t) + \frac{x^{\prime 2}+y^{\prime 2}}{2} (\Gamma t)^2)  \, ,
\label{eq:dbcs_rate_3770}
\end{eqnarray}
where $x^\prime = x  \mbox{cos}\delta + y \mbox{sin} \delta$ and
$y^\prime = -x \mbox{sin} \delta + y \mbox{cos} \delta$, and
$\delta$ is the strong phase between the DCS and CF amplitudes.
Recently,  a time-dependent analysis in $D \rightarrow K\pi$ has
been performed based on 384 fb$^{-1}$ luminosity at $\Upsilon(4S)$
by BaBar experiment~\cite{1}. By assuming $CP$ conservation, they
obtained the following neutral $D$ mixing results
\begin{eqnarray}
R_D &=& (3.03 \pm 0.16 \pm 0.10)\times 10^{-3} , \nonumber \\
x^{\prime^2}& =& (-0.22 \pm 0.30 \pm 0.21) \times 10^{-3},
\nonumber
\\
y^\prime &=& (9.7 \pm 4.4 \pm 3.1)\times 10^{-3}.
 \label{eq:kp_babar}
\end{eqnarray}
The result is inconsistent with the no-mixing hypothesis with a
significance of 3.9 standard deviations.  The results from BaBar
and Belle are in agreement within 2 standard deviation on the
exact analysis of $y^\prime$ measurement by using $D \rightarrow
K\pi$ as listed in Table~\ref{tab:experiments}.
\begin{table}[htbp]
  \caption{ Experimental results used in the paper. Only one error is quoted,
  we have combined in quadrature statistical and systematic contributions. }
  \center{\begin{tabular}{c|c|c|c} \hline\hline
  Parameter & BaBar ($\times 10^{-3}$)  & Belle($\times 10^{-3}$) &  Technique \\ \hline
  $x^{\prime^2}$ & -$0.22\pm 0.37$~\cite{1} &$0.18^{+0.21}_{-0.23}$~\cite{belle_kp_06} & $K\pi$ \\
  $y^{\prime}$ & $9.7\pm 5.4$~\cite{1} & $0.6^{+4.0}_{-3.9}$~\cite{belle_kp_06} & $K\pi$ \\
  $R_D$ & $3.03\pm 0.19$~\cite{1} & $3.64\pm 0.17$~\cite{belle_kp_06} & $K\pi$ \\
  $y_{CP}$ & - & $13.1\pm 4.1$~\cite{2} & $K^+K^-$, $\pi^+\pi^-$ \\
  $x$ & -& $8.0\pm3.4$~\cite{marko_belle_07} & $K_S \pi^+\pi^-$ \\
  $y$ & -& $3.3\pm 2.8$~\cite{marko_belle_07} & $K_S \pi^+\pi^-$ \\
  \hline \hline
 \end{tabular}
  \label{tab:experiments}}
\end{table}

At $\psi(3770)$ peak, to extract the mixing parameter $y$, one can
make use of rates for exclusive $D^0\Dzb$ combination, where both
the $D^0$ final states are specified (known as double tags or DT),
as well as inclusive rates, where either the $D^0$ or $\Dzb$ is
identified and the other $D^0$ decays generically (known as single
tags or ST)~\cite{asner_2005}.  With the DT tag
technique~\cite{markiii_1,markiii_2}, one can fully consider the
quantum correlation in $C=-1$ and $C=+1$ $D^0\Dzb$ pairs produced
in the reaction $e^+e^- \rightarrow D^0 \Dzb(n\pi^0)$ and $e^+e^-
\rightarrow D^0 \Dzb \gamma
(n\pi^0)$~\cite{bigi_tau,bigi_sanda,asner_2005}, respectively.

For the ST, in the limit of $CP$ conservation, the rate of $D^0$
decays into a $CP$ eigenstate is given as~\cite{asner_2005}:
\begin{eqnarray}
 \Gamma_{f_\eta}\equiv\Gamma(D^0 \rightarrow f_{\eta}) = 2A_{f_{\eta}}^2
\left[1-\eta  y \right],
 \label{eq:st_cp}
\end{eqnarray}
where $f_{\eta}$ is a $CP$ eigenstate with eigenvalue $\eta = \pm
1$, and $A_{f_{\eta}}= |\langle f_\eta | {\cal H}|D^0 \rangle|$ is
the real-valued decay amplitude.

For the DT case, Gronau {\it et. al.}~\cite{grossman_2001} and
Xing~\cite{xing_1997} have considered time-integrated decays into
correlated pairs of states, including the effects of non-zero
final state phase difference. As discussed in
Ref.~\cite{grossman_2001}, the rate of ($D^0 \Dzb)^{C=-1}
\rightarrow (l^\pm X)(f_\eta)$ is described
as~\cite{grossman_2001}:
\begin{eqnarray}
\Gamma_{l;f_\eta}\equiv \Gamma[(l^\pm X)(f_{\eta})] &= & A_{l^\pm
X}^2A_{f_\eta}^2 (1+ y^2) \nonumber \\
&\approx&  A_{l^\pm X}^2 A_{f_\eta}^2, \label{eq:dt_cp}
\end{eqnarray}
where $A_{l^\pm X} = | \langle l^\pm X|{\cal H}|D^0\rangle|$ is
real-valued amplitude for semileptonic decays, here, we neglect
$y^2$ term since $y\ll 1$.

For $C=-1$ initial $D^0\Dzb$ state,  $y$ can be expressed in term
of the ratios of DT rates and the double ratios of ST rates to DT
rates~\cite{asner_2005}:
\begin{eqnarray}
y = \frac{1}{4 } \left( \frac{\Gamma_{l; f_+}
\Gamma_{f_-}}{\Gamma_{l;f_-}\Gamma_{f_+}} -\frac{\Gamma_{l; f_-}
\Gamma_{f_+}}{\Gamma_{l;f_+}\Gamma_{f_-}} \right ).
\label{eq:dt_cp_y}
\end{eqnarray}
For a small $y$, its error, $\Delta (y)$, is approximately
$1/\sqrt{N_{l^\pm X}}$, where $N_{l^\pm X}$ is the total number of
$(l^\pm X)$ events tagged with $CP$-even and $CP$-odd eigenstates.
The number $N_{l^\pm X}$ of $CP$ tagged events is related to the
total number of $D^0 \Dzb$ pairs $N(D^0 \Dzb)$ through $N_{l^\pm
X} \approx N(D^0 \Dzb)[ {\cal B}(D^0 \rightarrow l^\pm +X)\times
{\cal B}(D^0 \rightarrow f_{\pm})\times \epsilon_{tag}] \approx
1.5\times 10^{-3} N(D^0 \Dzb)$, here we take the branching
ratio-times-efficiency factor (${\cal B}(D^0 \rightarrow
f_{\pm})\times \epsilon_{tag}$) for tagging $CP$ eigenstates is
about 1.1\% (the total branching ratio into $CP$ eigenstates is
larger than about 5\%~\cite{pdg_2006}). We find

\begin{eqnarray}
\Delta(y) = \frac{\pm 26}{\sqrt{N(D^0\Dzb)}} = \pm 0.003.
 \label{eq:dt_cp_dy}
\end{eqnarray}
If we take the central value of $y$ from the measurement of
$y_{CP}$ at Belle experiment~\cite{2},  thus, at BES-III
experiment, with 20$fb^{-1}$ data at $\psi(3770)$ peak, the
significance of the measurement of $y$ could be around 4.3
$\sigma$ deviation from zero~\cite{li_yang_d_2007}.

We can also take advantage of the coherence of the $D^0$ mesons
produced at the $\psi(3770)$ peak to extract the strong phase
difference $\delta$ between DCS and CF decay amplitudes. Because
the $CP$ properties of the final states produced in the decay of
the $\psi(3770)$ are anti-correlated~\cite{bigi_tau,bigi_sanda},
one $D^0$ state decaying into a final state with definite $CP$
properties immediately identifies or tags the $CP$ properties of
the other side.  As discussed in Ref.~\cite{grossman_2001}, the
process of one $D^0$ decaying to $K^-\pi^+$, while the other $D^0$
decaying to a $CP$ eigenstate $f_{\eta}$  can be described as
\begin{eqnarray}
\Gamma_{K\pi;f_\eta}\equiv \Gamma[(K^- \pi^+)(f_{\eta})] &\approx
& A^2A^2_{f_{\eta}}|1+ \eta
\sqrt{R_D} e^{-i \delta} |^2  \nonumber \\
&\approx&  A^2A^2_{f_{\eta}}(1+2 \eta \sqrt{R_D}
\mbox{cos}\delta),\nonumber \\
 \label{eq:besiii_delta_rD}
\end{eqnarray}
where $A = |\langle K^- \pi^+ |{\cal H}| D^0 \rangle |$ and
$A_{f_{\eta}} = |\langle f_{\eta} |{\cal H}| D^0 \rangle |$ are
the real-valued decay amplitudes, and we have neglected the $y^2$
terms in Eq.~(\ref{eq:besiii_delta_rD}). In order to estimate the
total sample of events needed to perform a useful measurement of
$\delta$,  one defined~\cite{grossman_2001,ian_2003} an asymmetry
\begin{eqnarray}
{\cal A} \equiv \frac{\Gamma_{K\pi;f_+} - \Gamma_{K\pi;
f_-}}{\Gamma_{K\pi;f_+} +\Gamma_{K\pi; f_-}},
\label{eq:besiii_delta_rD_a}
\end{eqnarray}
where $\Gamma_{K\pi;f_\pm}$ is defined in
Eq.~(\ref{eq:besiii_delta_rD}), which is the rates for the
$\psi(3770) \rightarrow D^0 \Dzb$ configuration to decay into
flavor eigenstates and a $CP$-eigenstates $f_\pm$.
Eq.~(\ref{eq:besiii_delta_rD}) implies a small asymmetry, ${\cal
A} = 2 \sqrt{R_D} \mbox{cos} \delta$. For a small asymmetry, a
general result is that its error $\Delta {\cal A}$ is
approximately $1/\sqrt{N_{K^-\pi^+}}$, where $N_{K^-\pi^+}$ is the
total number of events tagged with $CP$-even and $CP$-odd
eigenstates. Thus one obtained
\begin{eqnarray}
\Delta (\mbox{cos} \delta) \approx \frac{1}{2\sqrt{R_D}
\sqrt{N_{K^-\pi^+}}}. \label{eq:besiii_delta_rD_est}
\end{eqnarray}
The expected number $N_{K^-\pi^+}$ of $CP$-tagged events can be
connected to the total number of $D^0 \Dzb$ pairs $N(D^0 \Dzb)$
through $N_{K^-\pi^+} \approx N(D^0 \Dzb){\cal B}(D^0 \rightarrow
K^- \pi^+) \times {\cal B}(D^0 \rightarrow f_{\pm})\times
\epsilon_{tag} \approx 4.2 \times 10^{-4} N(D^0
\Dzb)$~\cite{grossman_2001}, here, as in Ref~\cite{grossman_2001},
we take the branching ratio-times-efficiency factor ${\cal B}(D^0
\rightarrow f_{\pm})\times \epsilon_{tag}=1.1\%$.  With the
measured $R_{D} = (3.03\pm 0.19)\times 10^{-3}$ and ${\cal B}(D^0
\rightarrow K^- \pi^+)=3.8\%$~\cite{pdg_2006}, one
found~\cite{grossman_2001}
\begin{eqnarray}
\Delta (\mbox{cos}\delta) \approx \frac{\pm 444}{\sqrt{N(D^0
\Dzb)}}. \label{eq:besiii_delta_rD_est_num}
\end{eqnarray}
At BES-III,  about $72 \times 10^6$ $D^0 \Dzb$ pairs can be
collected with 4 years' running. If considering both $K^- \pi^+$
and $K^+\pi^-$ final states,  we thus estimate that one may be
able to reach an accuracy of about 0.04 for cos$\delta$.

At BES-III, the measurement of $R_M$ can be performed
unambiguously with the following reactions~\cite{bigi_tau}:
\begin{eqnarray}
&(i)&\,\, e^+ e^- \rightarrow \psi(3770) \rightarrow D^0 \Dzb
\rightarrow
(K^\pm \pi^\mp)( K^\pm \pi^\mp), \nonumber \\
&(ii)&\, \, e^+ e^- \rightarrow \psi(3770) \rightarrow D^0 \Dzb
\rightarrow
(K^- e^+ \nu )(K^- e^+ \nu),  \nonumber \\
&(iii)& \, \, e^+e^- \rightarrow D^- D^{*+} \rightarrow (K^+ \pi^-
\pi^-) ( \pi^+_{soft} [K^+ e^- \nu]). \nonumber \\
\label{eq:besiii_rm_corr_1}
\end{eqnarray}
Reaction $(i)$ in Eq.~(\ref{eq:besiii_rm_corr_1}) can be
normalized to $D^0\Dzb \rightarrow (K^-\pi^+)(K^+\pi^-)$, the
following time-integrated ratio is obtained by neglecting $CP$
violation:
\begin{eqnarray}
\frac{N[(K^- \pi^+ )(K^-\pi^+)]}{N[(K^-\pi^+)(K^+\pi^-)]} \approx
\frac{x^2+y^2}{2} \equiv R_M. \label{eq:besiii_rm_measure_hadron}
\end{eqnarray}
For the case of semileptonic decay, as $(ii)$ in
Eq.~(\ref{eq:besiii_rm_corr_1}),  we have
\begin{eqnarray}
\frac{N(l^{\pm} l^\pm)}{N(l^\pm l^\mp)} = \frac{x^2+y^2}{2} \equiv
R_M, \label{eq:besiii_rm_measure}
\end{eqnarray}

In the limit of $CP$ conservation, by combing the measurements of
$x$ in $D^0 \rightarrow K_S \pi\pi$ and $y_{CP}$ from Belle,  one
can obtain $R_M =(1.18\pm 0.6)\times 10^{-4}$.  With 20fb$^{-1}$
data at BES-III, about 12 events for the precess $D^0\Dzb
\rightarrow (K^\pm \pi^\mp) (K^\pm \pi^\mp)$ can be produced. One
can observe 3.0 events after considering the selection efficiency
at BES-III, which could be about 25\% for the four charged
particles. The background contamination due to double particle
misidentification is about 0.6 event with 20$fb^{-1}$ data at
BES-III~\cite{kanglin_07}.

\subsection{$CP$ Violation in $D$ System}

For the direct $CP$ violation, the SM predictions are as large as
0.1\% for $D^0$ decays, and 1\% level for $D^+$ and $D_S$
decays~\cite{buccella}. At BES-III, one can also look at the $CP$
violation by exploiting the quantum coherence at the $\psi(3770)$.
Consider the case where both the $D^0$ and the $\Dzb$ decay into
$CP$ eigenstates, then the decays $\psi(3770) \rightarrow f^i_+
f^i_+$ or  $f^i_- f^i_-$ are forbidden, where $f_+$ ($f_-$)
denotes a $CP+$ eigenstate ($CP-$ eigenstate). This is because
$CP(f^i_\pm f^i_\pm) = CP(f^i_\pm)CP(f^i_\pm)(-1)^l = -1$,  while,
for the $l=1$ $\psi(3770)$ state, $CP(\psi(3770)) = +1$. Thus the
observation of a final state such as $(K^+K^-)(\pi^+\pi^-)$
constitutes evidence of the $CP$ violation. For
$(K^+K^-)(\pi^+\pi^-)$ mode, the sensitivity at BES-III is about
1\% level. Moreover, all pairs of the $CP$ eigenstates, where both
eigenstates are even or both are odd, can be summed over for the
$CP$ violation measurements at BES-III.

\section{Summary}

 In summary, the rare charm decays are searched for at CLEO-c. No
 evidence is found either for the rare (FCNC) decays
or for the forbidden (LNV) decays of charged $D$ mesons to
three-body final states with dielectrons.  The sensitivities of
the rare charm decays are  presented by assuming 20 fb$^{-1}$ at
$\psi(3770)$ peak with BES-III detector.  The measurements at
BES-III for these rare decays can be at the level of 10$^{-8}$.
The sensitivities for the neutral $D$ mixing and $CP$ violation
are also studied,  we find that  sensitivity for the lifetime
difference $y$ can be 0.003 with 20 fb$^{-1}$ data at $\psi(3770)$
peak, and the significance will be 4.3 $\sigma$ if the $y$ is at
1\% level.

\end{document}